\documentclass[11pt]{article}

\usepackage[dvips]{graphicx}
\usepackage{amsfonts}

\newcommand{\bb}{\begin{eqnarray}}
\newcommand{\ee}{\end{eqnarray}}
\newcommand{\beq}{\begin{equation}}
\newcommand{\eeq}{\end{equation}}
\newcommand{\ba}{\begin{array}}
\newcommand{\ea}{\end{array}}
\newcommand{\bd}{\begin{displaymath}}
\newcommand{\ed}{\end{displaymath}}

\def\l{{\lambda}}

\def\bbbn{{\mathbb N}}
\def\bbbc{{\mathbb C}}

\def\bbbz{{\mathbb Z}}

\def\bbbd{{\mathbb D}}

\def\eug{{\mathfrak g}}

\def\cL{{\cal L}}
\def\cR{{\cal R}}

\def\cF{{\cal F}}

\def\cp1{{\mathbb C\mathbb P}^1}

\def\dim{{\mbox{dim\,}}}
\begin{document}

\title{Reductions of integrable equations. Dihedral group}
\author{S Lombardo$^*$, A V Mikhailov$^{*+}$\\
$^*$ Department of Applied Mathematics, University of Leeds, Leeds LS2 
9JT, UK\\
$^+$ L.D.Landau Institute for Theoretical Physics, Chernogolovka, Russia (on leave)\\
{\small E-mail: sara@maths.leeds.ac.uk,
sashamik@maths.leeds.ac.uk}}

\date{}
\maketitle

\begin{abstract}
We discuss algebraic and analytic structure of rational Lax operators.
With algebraic reductions of Lax equations we associate a {\em reduction group} - a group of twisted automorphisms of the corresponding infinite dimensional Lie algebra. We present a complete study of dihedral
reductions for $sl(2,\bbbc)$ Lax operators with simple poles and corresponding integrable equations. In the last section we give three examples of dihedral reductions for $sl(N,\bbbc)$ Lax operators.   \\

\end{abstract}

\section{Introduction} 
The \emph{inverse spectral transform method} (IST) for integrating nonlinear 
partial differential equations (PDEs) is based on the \emph{Lax representation}
(see for example \cite{mnpz}, \cite{as}, \cite{cd}). In this setting, the 
nonlinear equation is equivalent to the compatibility condition of a pair 
of linear equations,  called \emph{Lax pair}. In general, given an integrable 
nonlinear equation, there is no algorithmic way to find its Lax 
representation.
Some progress in the problem of finding a Lax representation for a given 
nonlinear equation have been made by the Wahlquist and Estabrook method of 
pseudo-potentials and prolongation structures \cite{we1}, \cite{we2}, 
\cite{we3}. In some cases the Painlev\'e approach provides both 
integrability conditions and Lax representation (see for example 
\cite{fnt}). 

A different approach to this problem consists in starting from a quite 
general Lax representation which yields to a rather big (but integrable) 
system of equations. In general, such a system has too many degrees of 
freedom and is very complicated. However, it can contain smaller 
subsystems, relevant for the applications: many famous integrable 
equations (such as the sine-Gordon equation and the Tzetseica equation) 
are indeed the result of \emph{reductions} of more general integrable systems. 
This observation suggests to study and classify possible Lax 
representations and their reductions.

The first attempt to study algebraic reductions of Lax representations has 
been made in \cite{m1}, \cite{m2}, \cite{m3} and later in \cite{ggk}. The 
principal observation in the papers \cite{m1}, \cite{m2} was that  
reductions of integrable equations can be associated with a discrete 
symmetry group (called \emph{reduction group}) of the corresponding linear 
problems. The simplest example of such a symmetry is the conjugation for 
self-adjoint operators.

In this paper, which is meant to be the first of a series, we report new 
developments of this approach. The paper is organised as follows. In 
Section \ref{sec:laxr} we revisit the concept of Lax representation and 
its reductions. We define a quite general family of Lax operators  with 
rational dependence on the spectral parameter $\lambda$, on a simple
Lie algebra $\eug$. In Section 
\ref{sec:DN2x2} we study reductions of integrable equations corresponding 
to rational Lax pairs, with simple poles in $\l$ and $\eug=sl(2,\bbbc)$. 
Section \ref{sec:ir} concerns with the study of possible reductions a 
fairly general Lax pair whose underlying algebras is $\eug=sl(N,\bbbc)$. 
In both sections, reductions are obtained using realisations of the 
reduction group isomorphic to the dihedral group $\bbbd_N$. Some of the 
systems of equations obtained in these sections are new, to the 
best of our knowledge. Section \ref{sec:conclusions} is devoted to 
conclusions and remarks.

\section{Lax representation of integrable systems}\label{sec:laxr}
Let us start this preliminary section reviewing the concept of Lax 
representation of an integrable system. Here and afterwards subscripts 
mean partial differentiation.

\subsection{Example of a rational matrix Lax pair and gauge 
transformations}\label{sec:gauge}
Let us consider two linear equations
\beq\label{eq:pair}
L\psi=\psi_{x}-Q(x,t)\psi=0\, ,\quad M\psi=\psi_t-P(x,t)\psi=0,
\eeq
where $\psi,\,Q$ and $P$ are $N\times N$ matrix functions of $x$ and $t$. 
This is an over-determined system of equations and its compatibility 
condition
\begin{equation}\label{eq:trivpde}
 Q_t-P_x+[Q,P]=0
 \end{equation}
is equivalent to a system of nonlinear partial differential equations 
(PDEs). 
Its general solution is given by
\bd
Q=g_{x} g^{-1}\,  ,\quad P=g_{t} g^{-1}\, ,
\ed
where $g=g(x,t)$ is any nonsingular matrix function of $x$ and $t$. The 
same problem becomes nontrivial if one assumes that the matrices $Q$ and 
$P$ depend also on an auxiliary parameter $\lambda$, called \emph{spectral 
parameter}, and are polynomial or rational functions of this parameter. As 
a consequence, equation (\ref{eq:trivpde}) has to be satisfied for all 
values of $\lambda$ and this requirement leads to a nontrivial system of 
integrable, partial differential equations. In the latter case, one has to 
solve a certain matrix Riemann-Hilbert problem  to find a suitable matrix 
function $g(x,t,\lambda)$, which would provide the solution of the system 
(see for example \cite{mnpz}, \cite{zm}, \cite{zsh}). In order to receive 
a nontrivial set of nonlinear equations integrable by the IST method an 
{\em essential} dependence of $Q$ and $P$ in (\ref{eq:trivpde}) on the  
spectral parameter $\lambda$ is required, where essential dependence on 
$\lambda$ means that it cannot be  eliminated by a $\lambda$ dependent 
gauge transformation (see below).\\
As an example, let us consider the case of $Q$ and $P$ given by
\beq
Q=Q_0+\sum_{n=1}^{N_1}\frac{Q_n}{\lambda -a_n}\, ,\quad
P=P_0+\sum_{n=1}^{N_2}\frac{P_n}{\lambda -b_n}
\eeq
where it is assumed that the sets of complex constants 
$\{a_1,...,a_{N_1}|a_n\in\bbbc \}$ and $\{b_1,...,b_{N_2}|b_n\in \bbbc \}$ 
have empty
intersection and the matrix-coefficients $Q_{n},P_{n}$ may depend on
$x$ and $t$ only. Here and hereafter $\l\in\bbbc$. Under these assumptions the 
compatibility condition
(\ref{eq:trivpde}) leads to a system of $N_1+N_2 +1 $ equations
\begin{eqnarray}\label{eqrat0}
&& Q_{0t}-P_{0x}+[Q_0,P_0]=0\\
&& Q_{nt}+\big[ Q_n,
P_0+\sum_{m=1}^{N_2}\frac{P_m}{a_n-b_m}\big]=0 \label{eqratQ}\\
&& P_{nx}+\big[ P_n, Q_0-\sum_{m=1}^{N_1}\frac{Q_m}{a_n-b_m}\big]=0
\label{eqratP}
\end{eqnarray}
for $N_1 +N_2 +2 $ functions, i.e. the system obtained is 
under-determined.\\
Lax representations of this type and their reductions will be studied in 
Section \ref{sec:DN2x2}.

The system of equation (\ref{eqrat0})--(\ref{eqratP}) is invariant with
respect to transformations 
\begin{equation}\label{gauge1}
Q\to g^{-1} Q g-g^{-1}g_x\, ,\quad P\to g^{-1} P g-g^{-1}g_t\, ,
\end{equation}
where $g$ is any nonsingular matrix function of $x$ and $t$ (in
principle it may depend on the spectral parameter $\lambda$, but
in this section we shall assume $g=g(x,t)$). These transformations are 
called \emph{gauge transformations} \cite{zm}, \cite{zt}. In terms 
of matrices
$Q_n$ and $P_n$ the gauge transformation (\ref{gauge1}) takes the form
\begin{equation}\label{gauge2}
Q_0\to g^{-1} Q_0 g-g^{-1}g_x\, ,\quad
 P_0\to g^{-1} P_0 g-g^{-1}g_t\, ,\quad Q_n\to g^{-1} Q_n g
 \, ,\quad P_n\to g^{-1} P_n g\, ,
 \end{equation}
where $n\ge 1$.
There are several ways to fix the gauge. The natural or
\emph{canonical} choice \cite{zm} is to set $Q_0=P_0=0$. 
This is possible because the equations
\[ Q_0 g-g_x=0\, ,\quad P_0 g-g_t=0 \]
are compatible and their compatibility condition coincides with
(\ref{eqrat0}). In this gauge, the system becomes a well determined system
of $N_1+N_2$ matrix equations (\ref{eqratQ}), (\ref{eqratP}) for
$N_1+N_2$ matrix functions $Q_n,P_n$, $n\ge 1$. Another useful way to fix 
the gauge is to assume that the matrix $Q_1$ is
in a canonical (for instance diagonal) form. If the eigenvalues of
$Q_1$ are distinct, then the remaining  gauge freedom, i.e.
transformations (\ref{gauge2}) which do not change $Q_1$,
consists of nonsingular diagonal matrices $g$. This remaining
freedom can be used to make the matrix $Q_0$ off-diagonal. Such
choice of the gauge is called the \emph{pole} gauge \cite{zm} and
it also provides a well determined system of equations. 

The nature of the gauge transformations is obvious: the
compatibility condition (\ref{eq:trivpde}) is the commutativity condition
of two linear differential operators
\[ L=\partial_x-Q\, ,\quad M=\partial_t-P\, .\]
If the operators $L$ and $M$ commute ($[L,M]=0$), then the transformed 
operators
\begin{equation}\label{gLM}
\hat{L}=g^{-1}Lg\, ,\quad \hat{M}=g^{-1}Mg
\end{equation}
also commute. Transformations of the form
\begin{equation}\label{gLMA}
\hat{L}=-h^{-1}L^A h\, ,\quad \hat{M}=-h^{-1}M^A h\, , 
\end{equation}
where $L^A=-\partial_x+Q^{tr}$, $M^A=-\partial_x+P^{tr}$ stand for formally 
adjoint operators and ``tr'' stands for matrix transposition, are also gauge
transformations.
Gauge transformations form a group, the \emph{gauge group}. 

In the gauge transformations (\ref{gLM}), (\ref{gLMA}) the invertible 
matrices $g$, $h$ may depend on $x$, $t$, $\l$ or even be  differential or 
pseudo-differential matrix operators. The Miura and B\"acklund 
transformations can be viewed as special kind of gauge transformations.

\subsection{Lax representations and Lie algebras}\label{sec:sla}
Equations (\ref{eqrat0})--(\ref{eqratP}) are naturally defined on any
Lie algebra $\mathfrak{g}$. Indeed, if $Q_n,P_n\in\eug$, then all
commutators and derivatives also belong to $\eug$. 
Let $\eug$ be a finite dimensional Lie algebra over $\bbbc$.
It is easy to see that only simple algebras are related to
coupled nonlinear equations. Indeed, according to the Levi-Maltsev
theorem \cite{jacobson}, any finite dimensional Lie algebra $\eug$ over 
$\bbbc$ can be decomposed as a direct sum
\[ \eug={\cal R}\oplus {\cal S} \]
where ${\cal R}$ is a solvable radical of $\eug$ and ${\cal S}$ is
a semi-simple subalgebra. The semi-simple subalgebra ${\cal S}$, if
nontrivial, is a sum of simple subalgebras
\[ {\cal S}=\oplus {\cal S}_k \]
and the following commutation relations hold:
\[ [{\cal R},{\cal R}]\subset {\cal R}\, ,
\quad [{\cal S}_k,{\cal R}]\subset {\cal R}\, , \quad  [{\cal S}_k
{\cal S}_n ]= \delta_{k,n} {\cal S}_k \, .\] 
If we decompose $Q$ and
$P$ in (\ref{eq:trivpde}) according to the Levi-Maltsev decomposition 
theorem
\[ Q=Q_{\cal R}+\sum Q_{{\cal S}_k}\, ,\quad
P=P_{\cal R}+\sum P_{{\cal S}_k}\]
 we obviously receive
\begin{eqnarray}\label{sseq}
&& \partial _t Q_{{\cal S}_k}-\partial _x P_{{\cal
S}_k}+[Q_{{\cal S}_k},P_{{\cal S}_k}]=0\\
\label{req}
 &&\partial _t Q_{\cal R}-\partial _x P_{\cal R}+[Q_{\cal R},P_{\cal
 R}]+ [Q_{\cal R},\sum P_{{\cal S}_k}]+[\sum Q_{{\cal S}_k},P_{\cal
 R}]=0\, .
\end{eqnarray}
We see that equations (\ref{sseq}), corresponding to the
semi-simple part, can be treated separately from the rest of the
system and there is no coupling between equations corresponding
to different simple subalgebras. It is easy to show 
that, in the radical ${\cal R}$, one can choose a basis in which the
system of equations (\ref{req}) become ``triangular'', i.e. it is
an ordered chain of equations such that each equation in this
chain is a linear equation with respect to its own dependent variable and
with coefficients and an inhomogeneous part depending on
variables of the preceding equations and variables corresponding
to the semi-simple part. Therefore, the problem of integration of
the whole system is reduced to the integration of a closed system of the 
nonlinear equations
(\ref{sseq}) and then to the integration of a chain of linear
equations (\ref{req}) with variable coefficients.\\
In this paper we focus our attention on the study of nonlinear equations
and therefore we shall always assume that the underline algebra $\eug$
is simple (or semi-simple).

\subsection{General rational Lax operators}\label{sec:ia}
In this section we define a quite general family of Lax operators
with rational dependence on the spectral parameter $\lambda$. Let $\eug$
be a finite dimensional simple Lie algebra over $\bbbc$ and let 
$\{\eug_1,\ldots ,\eug_N\}$ be a basis of $\eug$ 
\begin{equation}\label{struc}
[\eug_i,\eug_j]=\sum_{r=1}^N C_{ij}^r\eug_r\, ,\qquad C_{ij}^r\in \bbbc\, .
\end{equation}

Let $X=\{ x_1,x_2, \ldots \}$ be a set of independent variables (in the 
previous example (\ref{eq:pair})  we used $x$ and $t$ variables).  With 
every variable $x_k$ we associate a
\emph{divisor of poles} $\Gamma_k = m_1 \cdot \gamma_1+\cdots + m_n \cdot 
\gamma_n$, i.e. a
finite set of points $\hat{\Gamma}_k=\{\gamma_{1},\ldots ,\gamma_{n}\}$ on 
the 
Riemann sphere together with their multiplicities $m_1,\ldots , m_n\in\bbbn$. 
Let $\cL (\Gamma_k)$ denote a linear space of rational functions of the 
spectral parameter $\lambda$ such that the only singularities of 
these functions are poles at the points of the divisor with multiplicity 
equal or less than the multiplicity of the point. The dimension of this 
linear space $\dim \cL (\Gamma _k)$ is $M_k=m_1+\cdots +m_n+1$. 
Let $e^k_1(\l),\,\ldots\, ,e^k_{M_k}(\l)$ be a basis  in $\cL (\Gamma _k)$.

With $x_k,\Gamma_k$ and $\eug$ we  associate the {\em general Lax 
operator}
\begin{equation}\label{laxoper} 
L_k =\frac{d}{dx_k}-\sum_{r=1}^N\sum_{\beta=1}^{M_k} \eug_r e^k_\beta (\l) 
u_{r\beta}^k\, ,
\end{equation}
where $u_{r\beta}^k$ are smooth functions of variables $X$ in a certain 
open domain, and if the set $X$ is infinite, we assume that every function 
depends on a finite number of variables only. A  set of such functions 
is denoted by $\cF$.

The Lax operator $L_k$ is parametrised by $N\times M_k$ 
functions $u_{r\beta}^k$.
Any two Lax operators $L_k,L_s$ form a Lax pair. The commutativity 
condition $[L_k,L_s]=0$
is equivalent to an integrable (by the spectral transform method) system of 
partial differential equations  for the functions 
$u_{r\beta}^k,u_{r'\beta'}^s\in\cF$. 

In order to write this system of PDEs in explicit form we introduce a 
basis $h^{ks}_\alpha (\l)\, ,\ \alpha =1,\ldots , M_k+M_s-1$ in $\cL 
(\Gamma_k+\Gamma_s)$ and expand elements  $e^k_\beta 
(\l),e^s_{\beta'}(\l)\in  \cL (\Gamma _k+\Gamma _s)$ and the products
 $e^k_\beta(\l) e^s_{\beta'}(\l)\in  \cL (\Gamma _k+\Gamma _s)$ in this 
new basis
\beq\label{strucr} \begin{array}{l}
e^k_\beta(\l)=\sum_{\alpha =1}^{M_k+M_s-1} F^{k\alpha}_\beta 
h^{ks}_\alpha(\l)\, ,\quad
e^s_\beta(\l)=\sum_{\alpha =1}^{M_k+M_s-1} F^{s\alpha}_\beta 
h^{ks}_\alpha(\l)\, \\ \\
e^k_\beta (\l)e^s_{\beta'}(\l) =\sum_{\alpha =1}^{M_k+M_s-1} 
H^{ks\alpha}_{\beta \beta'} h^{ks}_\alpha(\l)\, ,\qquad 
F^{k\alpha}_\beta,F^{s\alpha}_{\beta'},H^{ks\alpha}_{\beta 
\beta'}\in\bbbc\, .\end{array}
\eeq
Thus we have 
\begin{equation}\label{genkseq}
\sum_{\beta=1}^{M_k}F^{k\alpha}_\beta \frac{\partial 
u_{r\beta}^k}{\partial x_s}-
\sum_{\beta'=1}^{M_s}F^{s\alpha}_{\beta'} \frac{\partial 
u_{r\beta'}^s}{\partial x_k}+
\sum_{i,j=1}^{N}\sum_{\beta=1,\beta'=1}^{M_k,M_s} C_{ij}^r 
H^{ks\alpha}_{\beta \beta'}
u^k_{i\beta}u^s_{i\beta'}=0, 
\end{equation}
where $r=1,\ldots ,N$ and $\alpha=1,\ldots ,M_k+M_s-1$. 

Equation (\ref{genkseq}) is a system of nonlinear PDEs with constant 
coefficients. The system contains partial derivatives in variables $x_k$ 
and $x_s$ only, and if we assume that functions 
$u_{r\beta}^k,u_{r'\beta'}^s$ depend on other variables from the set $X$ 
we can treat these variables as parameters. The system obtained is {\em 
canonical}, it is uniquely defined by the choice of a simple Lie algebra 
$\eug$ and two divisors $\Gamma_k$ and $\Gamma_s$ (in some cases  the 
divisors may coincide, cf. the Lax pair for the $N$--wave equations 
\cite{mnpz}).
Systems corresponding to a different choice of the basis of $\eug, \cL 
(\Gamma _k), \cL (\Gamma _s)$ and $\cL (\Gamma _k+\Gamma _s)$ can be 
obtained from (\ref{genkseq}) by a linear invertible transformations of 
dependent variables.

The system of equations (\ref{genkseq})
is under-determined, it contains $N\times (M_k+M_s-1)$ equations on 
$N\times (M_k+M_s)$ dependent variables. Indeed, on the Riemann sphere, we 
have $\dim \cL (\Gamma _k+\Gamma_s)=\dim \cL (\Gamma _k)+\dim \cL (\Gamma 
_s)-1$ (if we consider a similar setup on an algebraic curve with nonzero 
genus, then the genus would contribute in the accounting for the 
dimensions of the corresponding linear spaces of rational functions on the 
curve \cite{zakmikcurve}). The difference between the number of dependent 
variables and the number of equations is equal to the dimension of the Lie 
algebra $\eug$. Fixing the gauge we finally obtain a well posed problem. 

If we have a family of Lax operators $L_1,\ldots ,L_k$ corresponding to 
divisors $\Gamma_1,\ldots ,\Gamma_k$, then the conditions $[L_i,L_j]=0\, , 
\ 1\le i<j\le k$ yield an over-determined system of nonlinear partial 
differential equations. A general solution and explicit exact partial 
solutions of this system can be constructed by the inverse transform 
method.

\subsection{The problem of reduction and the reduction group}\label{sec:rd}
The integrable system (\ref{genkseq}) obtained in the previous section is 
very general. It looks too big and may not resemble useful equations but,  
up to the best of our knowledge, all known $1+1$ dimensional systems of 
integrable partial differential equations are subsystems (reductions) of 
(\ref{genkseq}). The problem to find and classify all subsystems of a 
general system is known as the \emph{reduction problem}. 

Restrictions to subalgebras  $\hat{\eug}\subset\eug$ are obvious 
reductions of (\ref{genkseq}). For instance, in the system of equations 
(\ref{eqrat0})--(\ref{eqratP}) one can consider all matrices to be 
skew-symmetric and this is compatible with the dynamics since the 
commutator of skew-symmetric matrices is skew-symmetric. With this 
reduction we associate the automorphism $\phi :a\to -a^{tr}$ of the Lie 
algebra $sl(N,\bbbc)$. The set of all elements of $sl(N,\bbbc)$ which are 
invariant with respect to the automorphisms $\phi$ is obviously a 
subalgebra of skew-symmetric matrices $so(N)=\{ a\in sl(N,\bbbc)\, |\, 
a=\phi(a)\}$. A complete description of automorphisms of finite 
dimensional semi-simple Lie algebras is well
known (see for example \cite{jacobson}). In the case of $sl(N,\bbbc )$ 
automorphisms can be described as follows: for $sl(2,\bbbc )$ all 
automorphisms are inner, i.e. they can be represented as $\phi: a\to 
G^{-1}aG$, where $G\in SL(2,\bbbc)$.
For $sl(N,\bbbc )\, ,\ N>2$ the group of all automorphisms consists of 
inner $\phi: a\to G^{-1}aG$
and outer automorphisms $\psi: a\to -H^{-1}a^{tr}H$ ($G,H\in 
SL(N,\bbbc))$.

Let $\cR(\Gamma)$ be a ring of rational functions of variable $\l$ with 
poles at points of $\Gamma$ and regular elsewhere. Fractional-linear 
transformations of the complex plane $\l$, which map the divisor $\Gamma$ 
into itself, induce automorphisms of the ring $\cR(\Gamma)$. We denote 
 by $\mbox{Aut}\, \cR(\Gamma)$ the group of automorphisms of $\cR(\Gamma)$.
Any subgroup of $G_R\subset\mbox{Aut}\, \eug \times \mbox{Aut}\, \cR 
(\Gamma)$ is a group of automorphisms of $\eug (\Gamma 
)=\eug\bigotimes_\bbbc \cR(\Gamma)$. The group $G_R$  is called the {\em 
reduction group}. Elements of the reduction group can be viewed as {\em 
twisted} automorphisms, i.e. Lie algebra automorphisms and simultaneous 
fractional-linear transformations of the spectral parameter $\l$. For 
example, in the case $sl(N,\bbbc)$, the action of elements of the 
reduction group $G_R$ can be 
represented either as $a(\l)\to G^{-1}a(\sigma_G(\l))G$ or as $a(\l)\to 
-H^{-1}a^{tr}(\sigma_H(\l))H$, where $G,H\in SL(N,\bbbc)$ and 
$\sigma_G(\l)$, $\sigma_H(\l)$ are the corresponding fractional-linear 
transformations of the $\l$ plane.

The set 
$\eug _{G_R}(\Gamma)=\{ a\in  \eug (\Gamma)\, |\, a=\phi (a),\,\forall 
\phi\in G_R\}$ is a subalgebra of $\eug (\Gamma)$, which we shall call 
$G_R$--automorphic subalgebra. The restriction of the general Lax 
operator to the subalgebra $\eug _{G_R}(\Gamma)$ is a reduction,
with reduction group $G_R$. Restrictions to invariant subalgebras are 
equivalent to require some symmetry conditions for the Lax 
operators. For example, in the case $sl(N,\bbbc)$ they lead to symmetry 
conditions of the form
\begin{equation}\label{rgLaction}
L_k(\l)= G^{-1}L_k(\sigma_G(\l))G\, ,\qquad L_k(\l)= 
-H^{-1}L_k^{A}(\sigma_H(\l))H\, .
\end{equation}
Comparing with (\ref{gLM}) and (\ref{gLMA}), we see that the operators are 
invariant with respect to simultaneous gauge transformations\footnote{As 
we have already mentioned at the end of section \ref{sec:gauge} there are 
further generalisations when $G$ and $H$ are co-ordinate and $\l$  dependent or even (pseudo) differential operators.  We do not consider 
such generalised 
reductions in this paper.}
and fractional-linear transformations of the spectral parameter $\l$. The 
reduction group was originally introduced in \cite{m1}, \cite{m2}, our 
recent study of its algebraic structure and the corresponding automorphic Lie 
algebras will be published soon \cite{lm}.

\section{$\bbbd_{N}$-Reductions of Lax operators with simple poles, 
$\eug= sl(2,\bbbc)$.}\label{sec:DN2x2}
In this section we study dihedral  reductions of integrable
equations corresponding to rational Lax pairs with simple poles in 
the  spectral parameter $\l$ and $\eug=sl(2,\bbbc)$.

The dihedral group $\bbbd_{N}$ is the group of rotations and reflections of 
the
plane which preserve a regular polygon with $N$ vertices. It contains
$N$ rotations, which form a normal subgroup isomorphic to $\bbbz_{N}$ and 
$N$ reflections; its order is $2N$. If we denote by $s$ the rotation through 
an angle $2\pi /N$ and if $r$ is any of the reflections, then  each element of the
group can be
written uniquely either in the form $s^{k}$ or $rs^{k}$, $0 \leq k\leq 
N-1$.
In a more abstract way, the group $\bbbd_{N}$ can be defined as the group  
generated by two elements, $s$ and $r$, satisfying the identities
\beq
    s^{N}=r^{2}=id\, ,\quad r\,s\,r=s^{-1}\, .
    \label{eq:DN}
\eeq
In the case $N=2$ the group $\bbbd_2$ is abelian and isomorphic to 
$\bbbz_2 \times \bbbz_2$.

On the complex plane of spectral parameter $\lambda$ the group $\bbbd_{N}$ 
can be generated by 
two fractional-linear transformations
\begin{equation}\label{DNlambda} 
    \sigma _s: \lambda \to \omega \lambda\, 
,\quad \sigma_r :\lambda\to \lambda^{-1}\, ,\qquad
\omega=\exp \left(2 i \pi/N\right)\, . 
\end{equation}

If a divisor $\Gamma$ is invariant with respect to transformations 
(\ref{DNlambda}), it is a  union of a finite number of orbits of the group 
$\bbbd_N$.
The {\em orbit} of the group $\bbbd_N$ of a point $\gamma$ is defined as 
the  set of points
\[ D_N(\gamma)=\{\sigma_s^n(\gamma),\sigma_r\sigma_s^n(\gamma)\, |\, 
n=1,2,\ldots N\}\, .\]
When the point $\gamma$ is {\em generic}, i.e. $\gamma$ is not 
a fixed point of any (nontrivial) subgroup, the corresponding {\em 
generic} orbit has $2N$ points 
\begin{equation}
    \bbbd_N(\gamma)=\{\gamma,\,\omega\gamma,\,
    \omega^{2}\gamma,\ldots,\omega^{N-1}\gamma,\,\gamma^{-1},\,
    \omega\gamma^{-1},\, \omega^{2}\gamma^{-1},
    \ldots,\omega^{N-1}\gamma^{-1}\} .
    \label{eq:gODN}
\end{equation}
Fixed points of the group transformations (\ref{DNlambda}) belong to \emph{degenerated}
orbits. There exists a degenerated orbit with two elements
\begin{equation}
    \bbbd_N(0)=\{0,\infty\}\, ,
    \label{DN(0)}
\end{equation}
and two degenerated orbits with $N$ elements. For odd $N$ they are
\begin{equation}
    \bbbd_N(1)=\{\omega^n\, |\, n=1,2,\ldots ,N\}\, ,\,\, 
\bbbd_N(-1)=\{-\omega^n\, |\, n=1,2,\ldots ,N\}.
    \label{DNpm1}
\end{equation}
For even $N\,\,$  $\bbbd_N(1)=\bbbd_N(-1)$ and the second orbit with 
$N$ points is
\begin{equation}
    \bbbd_N(\omega^{1/2})=\{\omega^{n+1/2}\, |\, n=1,2,\ldots ,N\}\, .
    \label{DNeven}
\end{equation}
In the rest of this section we assume that $N$ is odd. 

With every orbit we associate a linear space of rational functions with 
simple poles at the points of the orbit. Let $\gamma$ be a generic point 
and $\bbbd_N(\gamma)$ the corresponding generic orbit. A natural way to 
construct a basis of rational functions in $\cL(\bbbd_N(\gamma))$ is to 
start with $e_0=1\, ,\,  e_{1}=1/(\lambda-\gamma)$
and apply all fractional-linear transformations of the group. As a 
result we receive a basis of $\cL(\bbbd_N(\gamma))$
\bd
e_{0}(\l)=1\, ,\quad e_k(\l)=\frac{1}{\omega^{k-1}\lambda-\gamma}\, 
,\quad e_{N+k}(\l)=\frac{1}{\lambda-\omega^{k-1}\gamma}\, ,\qquad 
k=1,2,\ldots ,N\, .
\ed 
This basis corresponds to the regular representation of the group 
$\bbbd_N$. A regular representation can be decomposed into a direct sum of 
irreducible representations. Such a decomposition suggests another natural 
and useful basis of $\cL(\bbbd_N(\gamma))$
\[
 E^\gamma_{0}=1\, 
,\quad
E^\gamma_{1}=\frac{\lambda^{N}+\gamma^{N}}{\lambda^{N}-\gamma^{N}}+\frac{1+\gamma^{N}\lambda^{N}}{1-\gamma^{N}\lambda^{N}}\, 
,\quad
E^\gamma_{2}=\frac{\lambda^{N}+\gamma^{N}}{\lambda^{N}-\gamma^{N}}-\frac{1+\gamma^{N}\lambda^{N}}{1-\gamma^{N}\lambda^{N}}\, 
,  \]
\[
E^\gamma_{4k-1}=\frac{\lambda^{k}}{\lambda^{N}-\gamma^{N}}\, 
,\quad E^\gamma_{4k}=\frac{\lambda^{N-k}}{1-\gamma^{N}\lambda^{N}}\, 
,\quad E^\gamma_{4k+1}=\frac{\lambda^{k}}{1-\gamma^{N}\lambda^{N}}\, 
,\quad E^\gamma_{4k+2}=\frac{\lambda^{N-k}}{\lambda^{N}-\gamma^{N}} .
\]
$k=1,\ldots,\frac{N-1}{2}$.

The general Lax operator corresponding to a generic orbit  $D_N(\gamma)$ 
is
\begin{equation}\label{lgam}
L_\gamma(\l) =\frac{d}{dx_\gamma}-\sum_{j=0}^{2N+1} E^\gamma_j(\l) 
U^\gamma _j
\end{equation}
where $U^{\gamma}_j$ are $2\times 2$ traceless matrix functions of 
independent variables (such as $x_\gamma$). Let $L_\mu(\l)$ be another 
Lax operator, corresponding to another generic orbit $D_N(\mu)$ and 
parametrised by matrix functions $U^\mu_j$. The condition 
$[L\gamma(\l),L_\mu(\l)]=0$ leads to a nonlinear system of $4N+1$ matrix 
partial differential equations. We will reduce this huge system to a 
simple, well determined system of six scalar partial differential equations.

Let $a(\l)\in sl(2,\bbbc)\bigotimes_\bbbc \cL(\bbbd_N(\gamma))$, i.e. 
\[ a(\l) = \sum_{j=0}^{2N+1} E^\gamma_j(\l) a _j\, ,\qquad a_j\in 
sl(2,\bbbc)\, ,\]
and consider the following linear transformations
\begin{equation} \label{rgroupdn}
g_s: a(\l)\to S(h)^{-1}a(\sigma_s(\l))S(h)\, ,\quad g_r: a(\l)\to 
R^{-1}a(\sigma_r(\l))R\, . \end{equation}
where $h$ is a fixed integer $\, 1\leq h\leq\frac{N-1}{2}$ that enumerates 
all different irreducible representations and 
\begin{equation}
    S=S(h)=\left(\begin{array}{cc}
    \omega^{-h} & 0 \\
    0 & \omega^{+h} \end{array} \right) \, ,\quad
    R=\left(\begin{array}{cc}
    0 & 1 \\
    1 & 0 \end{array} \right)\, .
    \label{eq:irrepr}
\end{equation}
The transformations $g_s$ and $g_r$ generate a linear representation of
$\bbbd_N$ in the 
linear space $sl(2,\bbbc)\bigotimes_\bbbc \cL(\bbbd_N(\gamma))$. It is 
easy to check that (\ref{rgroupdn}) generate a subgroup of the group of 
automorphisms of Lie algebra
$sl(2,\bbbc)\bigotimes\cR(\bbbd_N(\gamma))$. Let us consider invariant 
elements of $sl(2,\bbbc)\bigotimes_\bbbc \cL(\bbbd_N(\gamma))$, i.e. 
elements $a(\l)$ such that
\[ S(h)^{-1}a(\sigma_s(\l))S(h)=a(\l)\, ,\quad 
R^{-1}a(\sigma_r(\l))R=a(\l)\, .\]
For every given $h$ the subspace of invariant elements is three 
dimensional; a basis of this space can be written as 
\[\mathcal{E}^\gamma_1=E^\gamma_{ 2}\left(\begin{array}{rr}
    1 & 0 \\
    0 & -1 \end{array} \right)\, ,\quad 
\mathcal{E}^\gamma_2=\left(\begin{array}{cc}
    0 & E^\gamma_{3} \\
    E^\gamma_{4} & 0 \end{array} \right)\, ,\quad 
\mathcal{E}^\gamma_3=\left(\begin{array}{cc}
    0 & E^\gamma_{5} \\
    E^\gamma_{6} & 0 \end{array} \right)\, ,\]
where we fix $h=(N-1)/2$ since the final result does not depend 
on the choice of the representation.
Thus the reduced (invariant with respect to $\bbbd_N$ reduction group 
(\ref{rgroupdn})) Lax operator can be written as 
\begin{equation}\label{lgamred}
L_\gamma(\l) =\frac{d}{dx_\gamma}-q_1 \mathcal{E}^\gamma_1-q_2 
\mathcal{E}^\gamma_2-q_3 \mathcal{E}^\gamma_3\, ,
\end{equation}
where $q_1, q_2, q_3$ are scalar functions of independent variables. If 
\begin{equation}\label{lmured}
L_\mu(\l) =\frac{d}{dx_\mu}-p_1 \mathcal{E}^\mu_1-p_2 
\mathcal{E}^\mu_2-p_3 \mathcal{E}^\mu_3\, ,
\end{equation}
is the second operator in the Lax pair, corresponding to another generic orbit 
$D_N(\mu)$, then the condition $[L_\gamma,L_\mu]$ leads to a system of
integrable 
equations
\beq\label{eq:22}
\begin{array}{l}
q_{1t}+a(q_2p_3-p_2q_3)/2+b(q_2p_2-q_3p_3)/2=0\\
 q_{2t}-2\beta p_1q_2+4\gamma^N(ap_2+bp_3)q_1=0\\
q_{3t}+2\beta p_1q_3-4\gamma^N(ap_3+bp_2)q_1=0\\
p_{1x}+a(q_2p_3-p_2q_3)/2+b(q_2p_2-q_3p_3)/2=0\\
 p_{2x}+2\alpha q_1p_2-4\mu^N(aq_2-bq_3)p_1=0\\
p_{3x}-2\alpha q_1p_3+4\mu^N(aq_3-bq_2)p_1=0
\end{array}
\eeq
where $x=x_\gamma$, $t=x_\mu$ and $a$, $b$,  $\alpha$, $\beta$ are complex 
constants
\bd
a=(\gamma^N-\mu^N)^{-1}\, ,\quad b=(1-\gamma^N\mu^N)^{-1}\, ,
\ed
\bd 
\alpha=\frac{\gamma^{N}+\mu^{N}}{\gamma^{N}-\mu^{N}}+\frac{1+\gamma^{N}\mu^{N}}{1-\gamma^{N}\mu^{N}}\, 
, \quad 
\beta=\frac{\gamma^{N}+\mu^{N}}{\gamma^{N}-\mu^{N}}-\frac{1+\gamma^{N}\mu^{N}}{1-\gamma^{N}\mu^{N}}\, 
.
\ed
Clearly equations (\ref{eq:22}) can be rescaled and written in a different 
form, however they depend on parameters $\gamma^N$ and $\mu^N$, that 
cannot be removed by simple rescaling. What is surprising is that the 
resulting system does not depend on the choice of the representation. It 
does not depend on $N$ either! Indeed, equations 
corresponding to $N,\gamma,\mu$ coincide with equations corresponding to 
$N_1,\gamma_1,\mu_1$, 
provided the conditions $\gamma^N=\gamma_1^{N_1}$ and $\mu^N=\mu_1^{N_1}$ 
are satisfied. For the case of even $N$ we would receive equivalent 
equations, reflecting the fact that the corresponding automorphic Lie 
algebras are isomorphic \cite{lm}.

For the orbits $\bbbd_N(\pm 1)$  the bases of $\cL(\bbbd_N(\pm 1))$ in 
which the representation are decomposed in a direct sum of irreducible 
ones can be  written as
{\small \bd
F^\pm_{0}(\l)=1\, ,\,F^\pm _{1}(\l)=\frac{\lambda^{N}\pm 
1}{\lambda^{N}\mp 1}\, ,\,F^\pm_{2k}(\l)=\frac{\lambda^{k}}{\lambda^{N}\mp 
1}\, ,\,F^\pm_{2k+1}(\l)=\frac{\lambda^{N-k}}{1\mp \lambda^{N}}\, ,\quad 
k=1,\ldots,\frac{N-1}{2} \, .
\label{eq:B2}
\ed}
The corresponding $\bbbd_N$ invariant Lax operators are of the form 
\[ 
L_\pm(\l) =\frac{d}{dx_\pm}-\frac{1}{2}u^\pm_1 
F^\pm_1(\l)\left(\begin{array}{rr}
    1 & 0 \\
    0 & -1 \end{array} \right)-u^\pm_2 \left(\begin{array}{cc}
    0 & F^\pm_{2}(\l) \\
    F^\pm_{3}(\l) & 0 \end{array} \right)\, ,
\]
where again we assume $h=(N-1)/2$.
The  condition $[L_+,L_-]=0$ leads in this case to a rather simple system 
of equations
\begin{equation}\label{eq:1-1a}
    u^+_{2x_-}= u^+_{1}u^-_{2}\, ,\quad u^-_{2x_+}= u^-_{1}u^+_{2}\, 
,\quad u^+_{1x_-}=u^+_{2}u^-_{2}\, ,\quad u^-_{1x_+}=u^+_{2}u^-_{2}\, .
\end{equation}
It follows from (\ref{eq:1-1a}) that 
\bd
u^+_{2}u^+_{2x_-}=u^+_{1}u^+_{1x_-}\, ,\quad 
u^-_{2}u^-_{2x_+}=u^-_{1}u^-_{1x_+}\, ,
\ed
therefore we can partially integrate the equations and introduce the 
variables $\phi$ and $\theta$
\[  u^+_1=f(x_+) \cosh \phi\, ,\quad u^+_2=f(x_+) \sinh \phi\, ,\quad 
 u^-_1=g(x_-) \cosh \theta\, ,\quad u^-_2=g(x_-) \sinh \theta\, ,\]
where $f(x_+)$ and $g(x_-)$ are two arbitrary functions (which can be set 
to be equal to $1$ by a conformal transformation of the independent variables). In terms 
of $\phi$ and $\theta$ equations (\ref{eq:1-1a}) are
\beq\label{eq:1-1b} 
\phi _{x_-}=g(x_-)\sinh \theta\, ,\quad 
\theta_{x_+}=f(x_+)\sinh \phi \, ,
\eeq
they are nothing but a well known form of B\"acklund 
transformation for
the sinh-Gordon equation. Indeed, equations (\ref{eq:1-1b}) imply that
\beq\label{eq:1-1c}
\begin{array}{l}
(\theta+\phi)_{x_{+}x_{-}}=g(x_-)f(x_+)\sinh(\theta+\phi)\\
(\theta-\phi)_{x_{+}x_{-}}=g(x_-)f(x_+)\sinh(\theta-\phi)\, .
\end{array}
\eeq

Let us take $L_\gamma$ and $L_+$ as a Lax pair. The condition 
$[L_\gamma,L_+]=0$
is equivalent to the following system of equations
\begin{eqnarray}\label{g+1}
&&u^+_{1\, x_\gamma}=\frac{1}{\gamma^N-1}(q_2+q_3)u^+_2\\\label{g+2}
&&u^+_{2\, x_\gamma}=-\frac{2}{\gamma^N-1}(q_2+q_3)u^+_1\\\label{g+3}
&&q_{1\, x_+}=\frac{1}{2(\gamma^N-1)}(q_2+q_3) u^+_2\\\label{g+4}
&&q_{2\, x_+}=-\frac{4\gamma^N}{\gamma^N-1}q_1 
u^+_2+\frac{2(\gamma^N+1)}{\gamma^N-1}q_2 u^+_1\\\label{g+5}
&&q_{3\, x_+}=-\frac{4\gamma^N}{\gamma^N-1}q_1 
u^+_2-\frac{2(\gamma^N+1)}{\gamma^N-1}q_3 u^+_1\, .
\end{eqnarray}
It follows from (\ref{g+1}), (\ref{g+2}) that 
$2(u^+_1)^2+(u^+_2)^2=(f(x_+))^2$, where $f(x_+)$ is an arbitrary function 
and therefore 
\[ u^+_1=f(x_+)\cos \theta\, ,\quad u^+_2 =\sqrt{2}f(x_+)\sin \theta\, .\]
Now equations (\ref{g+1}), (\ref{g+2}) yield to
\[ q_2+q_3=\frac{1-\gamma^N}{\sqrt{2}}\theta_{x_\gamma}\, .\]
In variables
\[  \theta\, ,\quad u=2 q_1\, ,\quad 
v=-\frac{q_2-q_3}{\sqrt{2}(\gamma^N+1)}\]
equations (\ref{g+1})--(\ref{g+5}) take the form
\begin{equation}\label{thetauv}
\begin{array}{l}
u_{x_+}=f(x_+)(\cos \theta)_{x_\gamma}\\
v_{x_+}=f(x_+)(\sin \theta)_{x_\gamma}\\
\theta_{x_+\, x_\gamma}=\alpha f(x_+) u \sin \theta +\beta f(x_+) v \cos 
\theta
\end{array}
\end{equation}
where 
\[  \alpha =\frac{8\gamma^N}{(\gamma^N-1)^2}\, ,\quad \beta=4\left( 
\frac{\gamma^N+1}{\gamma^N-1}\right) ^2\, .\]
Changing the variable $x_+$ and rescaling $x_\gamma$ we can set
$ f(x_+)\to 1$, $\alpha \to \hat{\alpha}=\alpha/\beta$  and $\beta\to 1$.

The case of the orbit $\bbbd_N(0)=\{0,\infty\}$ yields a reducible $L_0$ 
operator 
\[ L_0=\frac{d}{dx_0}-p \left(\begin{array}{rr}
    0 & \l \\
    \l^{-1} & 0 \end{array}  \right) \, ,\quad h=\frac{N-1}{2}\, .\]
For other values of $h$ the operator is trivial.
It is not surprising, since we have imposed the condition that the poles 
are simple. The result become less trivial if we lift this condition (see 
\cite{lm}).

If we take $L_\gamma$ and $L_0$ as Lax pair, the condition 
$[L_\gamma,L_0]=0$
leads to 
\begin{equation}\label{g0eq}
 p_{x_\gamma}=4 q_1 p\, ,\quad 
q_{1\, x_0}=\frac{\gamma^N}{2}(q_3-q_2)\, ,\quad 
q_{2\, x_0}=-4\gamma^N q_1 p\, ,\quad 
q_{3\, x_0}=-4 q_1 p\, .\eeq
In the new variable $u=\log p$, after simple rescaling and a proper coordinate 
change, 
the system of equation (\ref{g0eq}) can be written as
\[ u_{x_\gamma\, x_0}=e^u \sqrt{1-(u_{x_\gamma})^2} .\]
The equation obtained is a well known example of the Liouville type 
equation \cite{zhs}. Considering $L_\pm$ and $L_0$ as Lax pair immediately leads to 
simple C-integrable system of equations.

For all Lax operators considered in this section the gauge freedom is 
completely fixed by the reduction group. Only scalar $\l$ independent 
gauge transformations commute with all elements of the reduction group.

Equations  (\ref{eq:22}) and (\ref{thetauv}) are \emph{new} 
integrable systems of equations, to the best of our knowledge. In 
particular, (\ref{thetauv}) may have interesting geometrical applications.

\section{Three examples of $ \bbbd_N$ reductions, 
$\eug=sl(N,\bbbc)$}\label{sec:ir}
Let us consider a fairly general Lax pair 
\begin{equation}\label{eq:L}
 L(x,t;\lambda)=\partial _x-X(x,t;\lambda)\, , \quad X=Q_{0}+Q\lambda
    +\bar{Q}\lambda^{-1}\, ,
\end{equation}
\begin{equation}\label{eq:M}
   M(x,t;\lambda)=\partial _t-T(x,t;\lambda)\, ,\quad T=P_{0}+P\lambda 
+\bar{P}\lambda^{-1}+
    Q^2\lambda^2 +\bar{Q}^2\lambda^{-2}\, ,
\end{equation}
with $Q_0,Q,\bar{Q},P_0,P,\bar{P}\in sl(N,\bbbc)$.
Note that $Q_{0}$ can be always set to zero by a gauge transformation. In 
this setting, the compatibility condition (\ref{eq:trivpde}) yields to the 
following set of equations
\begin{eqnarray}
\lambda^2 :\qquad && Q^{2}_{x}=[Q,P]\label{lam2}\nonumber\\
\lambda^1 :\qquad && 
Q_{t}-P_{x}+[Q,P_{0}]+[\bar{Q},Q^{2}]=0\label{lam1}\nonumber\\
\lambda^0 :\qquad && P_{0,x}=[Q,\bar{P}]+[\bar{Q},P]\label{lam0}\\
\lambda^{-1} :\qquad &&
\bar{Q}_{t}-\bar{P}_{x}+[\bar{Q},P_{0}]+[Q,\bar{Q}^{2}]=0
\label{lam-1}\nonumber\\
\lambda^{-2} :\qquad && \bar{Q}^{2}_{x}=[\bar{Q},\bar{P}]\, .
\label{lam-2}\nonumber
\end{eqnarray}
The system (\ref{lam0}) is a system of ($5(N^{2}-1)$) nonlinear coupled 
equations for the matrix entries. 
The group of automorphisms of Lie algebra $sl(N,\bbbc)$, $N\ge 3$ has 
both inner and outer automorphisms, allowing more possibilities for 
the realisation of the reduction group. Here we consider three different 
reductions of this system with reduction group isomorphic to $\bbbd_N$. 

\subsubsection{Case $1$: inner and outer automorphisms.}
Let us consider two transformations
\beq\label{eq:srexternal}
s\, :\quad L(\l)\mapsto S^{-1}\, L(\omega\l)\, S\, ,\quad r\, :\quad 
L(\l)\mapsto -L^{A}(1/\l)\, ,
\eeq
where $S$ is a $N\times N$ matrix given by 
$S_{ij}=\delta_{i,j}\omega^{N-i}$, with $\omega=e^{2i\pi/N}$, and where 
``$L^{A}$'' stands for formally adjoint operator $L^{A}=-\partial_{x}+X^{tr}$. 
Observe that, neglecting the spectral parameter $\l$, 
(\ref{eq:srexternal}) are nothing but two automorphisms of the algebras 
$sl(N,\bbbc)$; in particular, the first one is an inner  automorphism, 
while the second is outer \cite{jacobson}. They satisfy (\ref{eq:DN}) and 
therefore generate the dihedral group $\bbbd_N$.\\
Let us now require that both operators $L$ and $M$ are invariant under 
(\ref{eq:srexternal}). That leads to algebraic constraints on the matrices $X$ and $T$
\beq\label{eq:XTexternal}
\begin{array}{l}
X(\l)=S^{-1}\, X(\omega\l)\, S\, ,\quad X(\l)=-X^{tr}(1/\l)\, ,\\
T(\l)=S^{-1}\, T(\omega\l)\, S\, ,\quad T(\l)=-T^{tr}(1/\l)\, ,
\end{array}
\eeq
where ``tr'' stands for matrix transposition, which  imply
\begin{eqnarray}
&& P_{0}=0\, ,\nonumber\\
&&\bar{Q}=-Q^{tr}\, ,\quad \bar{P}=-P^{tr}\, ,\nonumber\\
&&Q=\mathbf{q}(x,t)\,\Delta \, ,\quad P=\mathbf{p}(x,t)\,\Delta\, 
,\nonumber
\end{eqnarray}
where $\mathbf{q}_{ij}=q_i(x,t) \delta_{i,j}\,$ and  $\mathbf{p}_{ij}=p_i 
(x,t)\delta_{i,j}$ are diagonal matrices and where $\Delta$ is the shift 
operator
$\Delta_{ij}=\delta_{i,j-1}$. Here and hereafter all indexes are counted modulo
$N$. As a 
consequence, (\ref{lam0}) reduces to the following system of 
$2(N-1)$ nonlinear coupled equations
\beq
\begin{array}{l}
Q_{t}-P_{x}+[\bar{Q},Q^{2}]=0\\\label{eq:a1}
Q_{x}^{2}=[Q\, ,\,P]\label{eq:a2}
\end{array}
\eeq
or, in components,
\begin{equation}
    q_{it}-p_{ix}+q_{i}q^{2}_{i+1}-q^{2}_{i-1}q_{i}=0
     \label{eq:qt1c}
\end{equation}
\begin{equation}\label{eq:qpcoupled}
    q_i p_{i+1}-p_i q_{i+1}-(q_i q_{i+1})_x=0\, .
\end{equation}
Solutions of (\ref{eq:qpcoupled}) can be parametrised by new variables 
$u_{i}$ and $v_{i}$ (similarly to \cite{m1}); indeed, let
\begin{equation}\label{uvvars}
    q_{i}=\exp(u_{i})\, ,\quad p_{i}=v_{i}\exp(u_{i})\, ,
\end{equation}
then
\begin{equation}\label{vvars}
v_{i}=-\sum_{r=1}^{N}\left(\frac{2\, \mbox{mod}\,(i-r-1,N)+1-N}{2N}
     \right)\left(u_{r}+u_{r+1}\right)_{x}\, .
    \label{eq:qt}
\end{equation}
In the new variables (\ref{eq:qt1c}) reads
\begin{equation}
u_{it}-v_{ix}-u_{ix}v_{i}+\exp(2u_{i+1})-\exp(2u_{i-1})=0\,,\,\,\,\,i=1,\ldots,N\, .
\end{equation}

In the case $N=3$ and $Q^{3}=I$ (or $\prod q_{i}=1$), equation 
(\ref{eq:a2}) can be solved explicitly 
\begin{equation}
   P=\frac{1}{3}[Q,QQ_{x}Q]\, .
     \label{eq:P3x3}
\end{equation}
In components
\bd
p_{1}=\frac{1}{3}q^{2}_{1}(q_{2}q_{3x}-q_{2x}q_{3})\, ,\quad
p_{2}=\frac{1}{3}q^{2}_{2}(q_{3}q_{1x}-q_{3x}q_{1})\, ,\quad
p_{3}=\frac{1}{3}q^{2}_{3}(q_{1}q_{2x}-q_{1x}q_{2})\, .
\ed
Hence, substituting $p_{i}$ into (\ref{eq:qt1c}) and rewriting
the equations in terms of $u_{i}$ variables we obtain
\beq\label{eq:equation1}
u_{1t}=\frac{1}{3}(u_{3xx}-u_{2xx})+\frac{1}{3}u_{1x}(u_{3x}-u_{2x})-\exp(2u_{2})
+\exp(2u_{3})
\eeq
and cyclic permutations of the indexes $1,\,2,\,3$.

\subsubsection{Case $2$: inner automorphisms.}
Let us now turn our attention to a different symmetry conditions
\beq\label{eq:internal}
L(\l)=S^{-1}\, L(\omega\l)\, S\, ,\quad L(\l)=R^{-1}\,L(1/\l)R
\eeq
where $R_{ij}=\delta _{i,N-j}$ (all indexes are counted modulo $N$).
Conditions (\ref{eq:internal}) are twisted inner automorphisms. In this 
case
\begin{eqnarray}
&& (P_{0})_{ij}=p_{0i}\delta_{i,j}\, ,\nonumber\\
&&\bar{Q}=R\,Q\,R\, ,\quad \bar{P}=R\,P\,R\, ,\nonumber\\
&&Q=\mathbf{q}(x,t)\,\Delta \, ,\quad P=\mathbf{p}(x,t)\,\Delta\, 
.\nonumber
\end{eqnarray}
Hence (\ref{lam0}) reduces to
\begin{eqnarray}
&& Q_{t}-P_{x}+[Q,P_{0}]+[\bar{Q},Q^{2}]=0\label{eq:a3}\\
&& P_{0x}=[Q,\bar{P}]+[\bar{Q},P]\label{eq:pxc}\\
&& Q_{x}^{2}=[Q\, ,\,P]\, ,
\end{eqnarray}
or, in components 
\begin{eqnarray}
&&
q_{it}=p_{ix}+p_{0i}q_{i}-q_{i}p_{0i+1}+q_{i}q_{i+1}q_{N-i-2}-q_{N-i}q_{i-1}q_{i}\label{eq:cl11}\\
&& p_{0ix}=q_{i}p_{N-i-1}-p_{i}q_{N-i-1}+q_{N-i}p_{i-1}
    -p_{N-i}q_{i-1}\\
&& q_i p_{i+1}-p_i q_{i+1}-(q_i q_{i+1})_x=0\,,\,\,\,\,i=1,\ldots,N\, 
.\label{eq:cl12}
\end{eqnarray}
For $N=3$ we can  use the result (\ref{eq:P3x3}) for $P$ to solve 
(\ref{eq:pxc}) and find
\begin{equation}
    P_{0}(x,t)=q_{2}(x,t)q_{3}(x,t)\, \mbox{diag}\, \{1\,,1\,,-2\}\, .
    \label{eq:matrixP_{0}}
\end{equation}
In  $u_{i}$ variables (\ref{uvvars}), we obtain
\beq
\begin{array}{l}
u_{1t}=\frac{1}{3}(u_{3xx}-u_{2xx})+\frac{1}{3}u_{1x}(u_{3x}-u_{2x})\\
u_{2t}=\frac{1}{3}(u_{1xx}-u_{3xx})+\frac{1}{3}u_{2x}(u_{1x}-u_{3x})
-\exp(2u_{1})+4\exp(u_{3}+u_{2})\\
u_{3t}=\frac{1}{3}(u_{2xx}-u_{1xx})+\frac{1}{3}u_{3x}(u_{2x}-u_{1x})
-4\exp(u_{3}+u_{2})+\exp(2u_{1})\, .
\end{array}
\label{eq:equation2}
\eeq

\subsubsection{Case $3$: symplectic automorphisms.}
In even dimensions $N=2n$  we can consider a third case given by
\begin{equation}\label{eq:symplectic}
    L(\lambda)=S^{-1}\,L(\omega\lambda)S\, ,\quad
    L(\lambda)=-J^{-1}\,L^{A}(1/\lambda)J\, ,
\end{equation}
where $J$ is the symplectic matrix $J={0\,\,\,I\choose -I\,0}$. 
From (\ref{eq:symplectic}) it follows that $P_{0}$ is a diagonal matrix
\bd
    P_{0}=\mbox{diag}\, \{p_{01}(x,t),\ldots 
,p_{0n}(x,t),-p_{01}(x,t),\ldots ,-p_{0n}(x,t)\}\, ,
\ed
while
\begin{equation}
    Q=\left(
    \begin{array}{ccccc}
      0  & q_{1} & 0 &  \ldots & 0 \\
      0 & 0 & q_{2}  & \ldots  & 0 \\
      0 & 0 & 0 & \ddots  & 0  \\
      0 & 0 & 0 & \ddots &  q_{2n-1} \\
      -q_{2n} & 0 & \ldots & 0 & 0
    \end{array}
    \right)\, ,
    \qquad
    P=\left(
     \begin{array}{ccccc}
      0  & p_{1} & 0 &  \ldots & 0 \\
      0 & 0 & p_{2}  & \ldots  & 0 \\
      0 & 0 & 0 & \ddots  & 0  \\
      0 & 0 & 0 & \ddots &  p_{2n-1} \\
      -p_{2n} & 0 & \ldots & 0 & 0
    \end{array}
    \right)\, ,
    \label{eq:QandP}\nonumber
\end{equation}
and
\bd
   \bar{Q}=-J^{-1}\,Q^{tr}J\, \quad \bar{P}=-J^{-1}\,P^{tr}J\, .
\ed
The equations for this case read
\begin{eqnarray}
&&
q_{it}=p_{ix}+q_{i}p_{0i}-p_{0i+1}q_{i}+q_{n+i-1}q_{i-1}q_{i}-q_{i}q_{i+1}q_{n+i+1}\label{eq:cl21}\\
&& p_{0ix}=-q_ip_{n+i}+q_{n+i}p_{i}+q_{i-1}p_{n+i-1}-q_{n+i-1}p_{i-1}\\
&& q_i p_{i+1}-p_i q_{i+1}-(q_i q_{i+1})_x=0\,,\,\,\,\,i=1,\ldots,2n\, 
.\label{eq:cl22}
\end{eqnarray}

For $n=2$ we have $[\bar{Q}\, ,\,Q^{2}]=0$ and equation (\ref{eq:cl21}) 
simplifies. In terms of $u_{i}$ variables (\ref{uvvars}) and (\ref{vvars}) the
system 
becomes
\beq\label{eq:equation3}
\begin{array}{l}
w_{1t}+\frac{1}{2}(w_{2x}w_{3x})=0\\
w_{2t}+w_{3xx}+\frac{1}{2}(w_{3x}w_{1x})-2l_{2}=0\\
w_{3t}-w_{2xx}+\frac{1}{2}(w_{2x}w_{1x})-2l_{1}=0
\end{array}
\eeq
\beq
l_{1x}=2w_{2x}\exp(-w_{1})\, ,\quad l_{2x}=-2w_{3x}\exp(w_{1})\, ,  
\eeq
where $w_{1}=u_{1}+u_{3}$, $w_{2}=u_{1}-u_{3}$, 
$w_{3}=u_{2}-u_{4}$ and where $l_{1}=({p}_{01}+p_{02})$,
$l_{2}=({p}_{01}-p_{02})$.

System (\ref{eq:equation3}) is \emph{new}, 
to the best of our knowledge. Moreover, for $N$ general all systems of 
equations obtained in this sections can be regarded as new examples of 
lattice equations.

\section{Conclusions and remarks}\label{sec:conclusions}
All properties of integrable equations are encoded in their Lax
representations. Therefore, the description of the variety of 
integrable equations and their reductions can be pursued starting from the
theory of Lax operators. In particular, the problem of reductions can be studied
imposing symmetry conditions on the Lax operators, reducing the corresponding
nonlinear systems of equations to smaller subsystems. 
Symmetries of Lax representations form a group - the reduction group 
\cite{m1}, \cite{m2}, \cite{m3}, \cite{lm}. The reduction group approach has
been used 
to describe Lie-algebraic reductions of the $N$--wave equation \cite{ggk}, 
to find Lax representation for a number of new Nonlinear Schr\"odinger 
type equations \cite{msy} and to build explicit solutions for the 
Landau-Lifschits equation \cite{mll}. In this paper we have revisited the 
reduction group approach, illustrated it by a number of examples and 
motivated our further study. There are several natural directions for 
development.

The equations obtained in the previous section are integrable for any $N$. 
Assuming $N\to\infty$ and taking continuous limits one can find $2+1$ 
dimensional integrable equations (and corresponding Lax representations 
-- all structures, such as symmetries, conservation laws, etc. can be 
recomputed in these limits). There are several ways to take a continuous 
limit, the result depend on the balance of non-linearity and dispersion. 
For example, the Kadomtsev-Petviashvili equation 
\begin{equation}
u_T=\frac{1}{3} u_{yyy}-D_y^{-1}(u_{XX})+6 uu_y\, . 
    \label{eq:kp}
\end{equation}
can be recovered from (\ref{eq:qt1c})--(\ref{eq:qpcoupled}) as a 
continuous limit if we assume $q=\exp(h^{2}u)$, where $h$ is the 
lattice step, and perform a Galilean transformation. On the other 
side, performing rather 
na\"ive expansions $q_{i\pm1}=q\pm h q_y+O(h^2) $ and $p_{i\pm1}=p\pm h 
p_y+O(h^2) $, 
we would receive, after proper rescaling, the hydrodynamic type equation 
\begin{equation}
    u_t=D_y\{(D_{y}^{-1}D_x)^{2}\, 
u+\frac{1}{2}(D_y^{-1}D_x\, u)^{2}+e^{(2u)}\}\, .
    \label{eq:slq}
\end{equation}

Similarly, equations (\ref{eq:cl11})--(\ref{eq:cl12}) yield to the 
following systems of equations in $2+1$ dimensions 
\begin{equation}
\begin{array}{l}
u_t=D_y\{(D_{y}^{-1}D_x)^{2}\, 
u+\frac{1}{2}(D_y^{-1}D_x\,
u)^{2}+D_{x}^{-1}D_y[e^{(u+v)}\,D_y^{-1}D_x(u+v)]+e^{(u+v)}\}\\
v_t=-D_y\{(D_{y}^{-1}D_x)^{2}\, 
v+\frac{1}{2}(D_y^{-1}D_x\,
v)^{2}+D_{x}^{-1}D_y[e^{(u+v)}\,D_y^{-1}D_x(u+v)]+e^{(u+v)}\}\, , 
\end{array}
    \label{eq:conlim3}
\end{equation}
while and (\ref{eq:cl21})--(\ref{eq:cl22}) yield to
\begin{equation}
\begin{array}{l}
u_t=D_y\{(D_{y}^{-1}D_x)^{2}\, 
u+\frac{1}{2}(D_y^{-1}D_x\,
u)^{2}+D_{x}^{-1}D_y[e^{(u+v)}\,D_y^{-1}D_x(u-v)]+e^{(u+v)}\}\\
v_t=D_y\{(D_{y}^{-1}D_x)^{2}\, 
v+\frac{1}{2}(D_y^{-1}D_x\,
v)^{2}+D_{x}^{-1}D_y[e^{(u+v)}\,D_y^{-1}D_x(u-v)]+e^{(u+v)}\}\, . 
\end{array} 
    \label{eq:conlim4}
\end{equation}
Notice that (\ref{eq:slq}) can be obtained from (\ref{eq:conlim4}) by setting 
$v=u$.

A simple modification of the procedure enables us to study Lax operators with
non commutative
(matrix) entries. For example, we can treat elements of  $sl(2M,\bbbc)$, as 
$2\times 2$ matrices with $M\times M$ matrix entries. The corresponding 
reduced systems can be viewed as a system of equations for 
non abelian variables. The following system of equations 
\beq\label{eq:nonabelian}
\begin{array}{l}
\mathbf{q}_{0t}-\mathbf{p}_{0x}+[\mathbf{q}_{0}\, 
,\,\mathbf{p}_{0}]+[\mathbf{q}_{1}\, ,\,\mathbf{p}_{1}]=0\\
\mathbf{q}_{1t}+[\mathbf{q}_{1}\, ,\,\mathbf{p}_{0}]+\{\mathbf{q}_{2}\, 
,\,\mathbf{p}_{2}\}/4=0\\
\mathbf{q}_{2t}+[\mathbf{q}_{2}\, ,\,\mathbf{p}_{0}]+\{\mathbf{q}_{1}\, 
,\,\mathbf{p}_{2}\}=0\\
\mathbf{p}_{1x}+[\mathbf{p}_{1}\, ,\,\mathbf{q}_{0}]+\{\mathbf{q}_{2}\, 
,\,\mathbf{p}_{2}\}/4=0\\
\mathbf{p}_{2x}+[\mathbf{p}_{2}\, ,\,\mathbf{q}_{0}]+\{\mathbf{q}_{2}\, 
,\,\mathbf{p}_{1}\}=0
\end{array}
\eeq
where the variables $\mathbf{q}_{i}$, $\mathbf{p}_{i}$ are elements of a 
non commutative free algebra (or matrices of any size) and 
$\{\mathbf{q}_{i}\, 
,\,\mathbf{p}_{j}\}=\mathbf{q}_{i}\mathbf{p}_{j}+\mathbf{p}_{j}\mathbf{q}_{i}$,
is a nonabelian generalisation of (\ref{eq:1-1b}) (the B\"acklund 
transformations for the sinh-Gordon equation). In the non abelian case we 
have to fix the gauge freedom further in order to make equations (\ref{eq:nonabelian}) well determined.

A very important issue is the solution of nonlinear integrable models. The
reduction group 
proves to be not only a very useful tool to find \emph{new} integrable 
equations 
and classify Lax pairs but also a necessary instrument in this 
context. Indeed, without the constraints imposed by the reduction group on 
the 
spectral data it is not even possible to formulate the inverse problem, 
which 
would lead to explicit solutions of the integrable equation. The general set up 
of the 
correspondence between reduction groups and analyticity properties of the 
spectral data is one of the next issues on our research agenda. 

Another challenging problem is the study of automorphic Lie algebras in 
a pure algebraic way. They can be always linked back to Lax operators, 
Baxter's $R$--matrix equations, etc. Moreover, the problem of a 
complete description of rational automorphic Lie algebras seems to be 
feasible. Our optimism is based upon a simple group-theoretical observation 
\cite{lm} and a remarkable theorem of Felix Klein 
\cite{klein}: {\em the complete list of finite groups of fractional-linear 
transformations is given by the cyclic group $\bbbz_{N}$, the dihedral group 
$\bbbd_{N}$ and the groups of symmetry of Plato solids, i.e. 
the tetrahedral group, the octahedral group and the icosahedral group}. It would
be interesting to generalise the theory of automorphic 
Lie algebras to the cases of elliptic and higher genus algebraic curves. We 
believe that automorphic Lie algebras will find applications far beyond 
the theory of Lax operators.

\noindent\textbf{Acknowledgments}\\
We would like to thank the Newton Institute for Mathematical Sciences 
where we started to work on this paper. The initial stage of the work of 
S L was supported by the University of 
Leeds \emph{William Wright Smith} scholarship and successively by a grant of
the Swedish foundation \emph{Blanceflor 
Boncompagni-Ludovisi, 
n\`ee Bildt}, for which S L is most grateful. The work of A M was 
partially supported by RFBR grant 02-01-00431.

%%%%%%%%%%%%%%%%%%%%%%%%%%%%%%%%%%%%%%%%%%%%%%%%%%%%%%%%%%%%%%%%%%%%%%%%

%%%%%%%%%%%%%%%%%%%%%%%%%%%%%%%%%%%%%%%%%%%%%%%%%%%%%%%%%%%%%%%%%%%%%%%%

\end{document}